\documentclass[lettersize,journal]{IEEEtran}
\usepackage{multirow}
\usepackage{algpseudocode} 
\usepackage{amsmath}
\usepackage{threeparttable}
\usepackage{booktabs}
\usepackage{tabularx}
\usepackage{amsmath,amsfonts}
\usepackage{xcolor}
\usepackage{algorithm}
\usepackage{array}
\usepackage[caption=false,font=normalsize,labelfont=sf,textfont=sf]{subfig}
\usepackage{textcomp}
\usepackage{stfloats}
\usepackage{url}
\usepackage{verbatim}
\usepackage{graphicx}
\usepackage{cite}
\usepackage{amsmath}
\usepackage{algorithm}

\usepackage{enumerate}
\usepackage{makecell}
\usepackage{fancyhdr}

\newcommand{\Ysys}{Y^{\text{sys}}}
\newcommand{\Zsys}{Z^{\text{sys}}}

\newcommand{\dq}{d\text{-}q}
\newcommand{\Ys}{\boldsymbol{Y}}
\newcommand{\OPS}{\mathcal{O}}
\newcommand{\pu}{\textit{p.u.}}

\begin{document}
\title{A Physics-Informed Neural Network for Small-\\Signal Stability in Multi-Inverter Power Systems }


\author{Hanxi~Chen, Xiangyu~Meng, \IEEEmembership{Member, IEEE}, Jianhong~Wang, Yue~Zhu, \IEEEmembership{Member, IEEE} 
}

\maketitle




\thispagestyle{fancy}
\lhead{IEEE TRANSACTIONS ON POWER SYSTEMS}
\rhead{\thepage}
\cfoot{}
\renewcommand{\headrulewidth}{0pt}
\pagestyle{fancy}

\begin{abstract}
The whole-system impedance model has proven a powerful tool for assessing the small-signal stability of multi-inverter power systems; however, its application is limited to a small range around a steady-state operating point due to the inherent assumptions of time invariance and linearisation. In this paper, a dedicated physics-informed neural network (PINN) for small-signal stability analysis in high-dimensional multi-inverter power systems is developed. The PINN is trained with step-response data produced from limited sets of system electromagnetic transient (EMT) simulations, and the trained model can predict the poles and residues of the whole-system impedance/admittance model, i.e., the transfer functions, across the full operating space. Such a PINN offers unique insights into system stability that surpass what conventional analytical methods or EMT simulations can achieve. By characterising how the impedance model evolves with power flow variations, it predicts the dynamic behaviour of the time-varying system and reveals oscillation risks that may emerge while identifying their root causes. It also provides direct visualisation of the possible range of oscillatory modes under a given power flow condition, enabling an optimal generation distribution while maintaining safe operation of the system. The proposed PINN is fully validated on a 2-IBR system and a 4-IBR system, with its application details presented.
\end{abstract}

\begin{IEEEkeywords}
Physics-informed neural network, impedance prediction, small-signal stability, multi-inverter power systems
\end{IEEEkeywords}

\section{Introduction}
With the increasing integration of renewable energy, power systems are undergoing a transition from being dominated by synchronous generators (SGs) to those dominated by inverter-based resources (IBRs)\cite{Hatziargyriou2021Definition,Blaabjerg2023Power}. This transformation significantly alters the system dynamics and has reshaped the approaches to stability analysis, particularly the small-signal stability \cite{Wang2019Harmonic,Gu2023Power, Cheng2023Real}. The conventional state-space model \cite{Singh2013IEEE}, established from differential equations and proven effective for small-signal stability analysis in SG-dominated systems, becomes inappropriate in IBR-dominated systems for several reasons. First, as IBR control schemes grow increasingly complex, for example, with the adoption of model predictive control\cite{Hu2015Model}, it becomes extremely difficult to formulate the differential equations of IBRs analytically. Second, IBR vendors treat control designs as intellectual property and therefore do not disclose the necessary control algorithms and parameters to system operators. These factors impede effective modelling of the system for small-signal stability assessment.

By far, the widely adopted approach for stability assessment in multi-inverter power systems is electromagnetic transient (EMT) simulations, as system operators typically require IBR vendors to provide encrypted black-box EMT models. The EMT simulations provide detailed waveforms that directly show the system's response to perturbations. However, EMT studies are computationally intensive and often time-consuming. More importantly, they offer only a descriptive view of system behaviour, without revealing the underlying mechanisms or root causes of instability. Given such restrictions, recent studies have established impedance model, or equivalently, admittance model, for small-signal stability analysis in multi-inverter systems. Such models are networked impedance models represented in matrix form, referred to as the whole-system impedance model \cite{Gu2021Impedance, Zhang2020Impedance}. Eigenvalue analysis and participation factors have been developed in whole-system impedance model \cite{Zhu2022RootCause, Liao2022Frequency-Domain, Jiang2024Impedance}. A systematic impedance-based participation analysis approach, known as the Grey-box approach, was proposed in \cite{Zhu2022Participation} and proves that the poles of whole-system impedance are the system eigenvalues, while the corresponding residues represent impedance participation factors. These developments have established the fundamentals of analysing the small-signal stability of multi-inverter systems using the impedance model in a manner similar to the conventional state-space model. 




As a black-box model represented by a set of frequency spectra or transfer functions, the impedance model captures the small-signal dynamics of the entire network while masking the design details. Moreover, it can be measured via frequency scanning (FS) in EMT simulations at differrent frequency points (FPs), regardless of the complexity of the IBR controller. Despite the advantages, the practical use of impedance-based stability analysis in power systems faces several critical challenges. First, there is no industrial standard for IBR vendors to provide impedance models. Second, FS for impedance measurement in the EMT simulation is also time-consuming, especially for multi-inverter systems where FS is required at each bus with an IBR connected. More importantly, the impedance-based analysis is only valid around a specific steady-state operating point (OP) set by the power flow. Once the OP changes, the impedance model changes accordingly, and FS must be repeated \cite{Ramakrishna2024Stability,Gong2021DQ}.
In a multi-inverter system with a high-dimensional OP space, it is essentially impractical to conduct FS for all possible operating points and across all buses.



Given the strong dependence of impedance models on OPs, machine learning (ML) based impedance prediction methods have been proposed in recent years \cite{Zhang2021Artificial, Zhang2023Transfer, Li2024Machine, Liao2024Neural, Mohammed2024Support,Li2025Sequence,Lyu2024Data, Wu2025Impedance}, with the aim of predicting the impedance spectra at an arbitrary OP. Most studies focus on a single-inverter system and predict the discretised impedance spectra in the $\dq$ frame. In \cite{Zhang2021Artificial}, a feedforward neural network (FNN) is developed for impedance prediction of an inverter at different OPs. The FNN is trained using frequency response data (FRD) obtained from FS conducted at multiple OPs, enabling the prediction of impedance spectra at unseen OPs. Although the prediction is limited to a one-dimensional OP space, where only $I_d$ varies while $V_d$, $V_q$, and $I_q$ are fixed, it serves as a clear proof of concept that ML can achieve accurate impedance prediction. Further efforts have been devoted to reducing the burden of training data collection and extending prediction capability to higher-dimensional operating spaces. The related work is summarised and compared in Table \ref{tab:method_comparison}.

\begin{table*}[!t]
\centering
\begin{threeparttable}
\caption{Comparisons of ML-Based Impedance Prediction Methods in Inverter-Dominated Systems}\vspace{-0.1cm}
\label{tab:method_comparison}

\setlength{\tabcolsep}{3.5pt}
\scriptsize
\renewcommand{\arraystretch}{1.35}
\begin{tabular}{p{2.0cm}| p{4cm}| p{1.2cm}| p{1.2cm} |p{1.3cm} |p{6.6cm}}
\toprule
\textbf{Machine Learning Algorithm}
& \textbf{Predicted Results}
& \textbf{OP Order}
& \textbf{Training Data Type}
& \textbf{Training Data Size}
& \textbf{Key Features} 
\\
\hline

Transfer learning (TL) \cite{Zhang2023Transfer}
& Spectra of harmonic impedance model of a MMC system
& 1 ($I_{out}$)
& FRD
& 15~OPs $\times$ 20~FPs
& Training an offline model, and feeding the model with data from online frequency scanning using TL for online identification
\\
\hline
Feedforward neural network (FNN) \cite{Li2024Machine}
&Spectra of $\dq$ frame impedance model of a grid-connected inverter
&3 $(P,Q,V)$
&FRD
&1084~OPs $\times$ 20~FPs
& Demonstrating an end-to-end ML framework for impedance predictions of a grid-connected inverter.\\
\hline
FNN \cite{Liao2024Neural}
& Spectra of $\dq$ frame impedance model of a grid-connected inverter
& 3 $(P,Q,V)$
& FRD
& 576~OPs $\times$ 100~FPs
& Using the number of poles and zeros of impedance model as latent features for a multi-layer FNN to reduce training data size\\ \hline

Support vector machine (SVM) \cite{Mohammed2024Support}
& Spectra of $\dq$ frame impedance model of a grid-connected inverter or 2 identical parallel inverters
& 3 $(P,Q,V)$
& FRD
& 18~OPs $\times$ 30~FPs
& Applying an efficient SVM based approach that demands only a small OP dataset for training\\ \hline

Bidirectional long short-term memory (Bi-LSTM) \cite{Li2025Sequence}
& Spectra of sequence frame impedance model of a grid-connected inverter
& 1 $(I_{out})$
& FRD
& 15~OPs $\times$ 83~FPs
& Using Bi-LSTM for higher accuracy of predictions compared with neural networks\\ \hline

Artifical neural network (ANN) \cite{Lyu2024Data}
& Spectra of $\dq$ frame impedance model of a wind turbine generator and an MMC
& 1 $(P_{out})$
& FRD
& 10~OPs $\times$ unknown FPs 
& Training ANNs offline for impedance prediction of each device, which are then aggregated for online stability assessment of a multi-inverter system \\ \hline

Stacked auto encoder (SAE) \cite{Wu2025Impedance}
& Spectra of the aggregated $\dq$ frame impedance model of 2 different grid-connected inverters
& 5 ($V_{PCC}$, $I_{d1}$, $I_{q1}$, $I_{d2}$, $I_{q2}$)
& FRD
& 4095~OPs $\times$ 89~FPs
& Using features extracted from impedance profiles to replace OPs for impedance characterisation for higher-order OP space\\ \hline

\textcolor{red}{
PINN (this work)}
& \textcolor{red}{Poles and residues of the $\dq$ frame whole-system admittance model of a meshed network with 1 SG and 4 different inverters connected}
& \textcolor{red}{11, defined in \eqref{eq_OP_5bus}}
& \textcolor{red}{SRD} 
& \textcolor{red}{1000~OPs $\times$ 1~s window length}
& \textcolor{red}{Using SRD to reduce training costs, and employing Sobol-sequence-based sampling strategy for high-order OP space}\\

\bottomrule
\end{tabular}
\end{threeparttable}
\vspace{-0.3cm}
\end{table*}

The use of ML-based impedance prediction methods has shown promise for single-inverter connection studies. However, their extensions to multi-inverter power systems face some critical challenges. First, the dimension of OP space expands exponentially with the number of IBRs. To ensure accurate predictions at arbitrary OPs, the training data must be collected from a large number of OPs to ensure a sufficient coverage of the high-dimensional space, which is apparently impractical when using FS, since each round of FS is very time-consuming. Second, while the predicted frequency spectra are informative for single-inverter studies, where phase margin and crossover frequency are of interest, they offer limited interpretability in multi-inverter systems. In such multi-input-multi-output (MIMO) systems, the poles and residues of transfer functions provide more meaningful insights \cite{Zhu2022RootCause}.


To address the aforementioned challenges, this paper develops a modular PINN specifically for predicting whole-system impedance model in multi-inverter power systems. Compared with existing ML-based methods, this PINN features the following key advancements: (i) instead of being trained with FRD from the time-consuming FS, the proposed PINN is trained using time-domain step-response data (SRD) with a 1~s window, therefore greatly reducing the time costs for data collection in high-dimensional OP spaces. (ii) Instead of predicting the frequency spectra, the PINN predicts the poles and residues of the whole-system impedance model, i.e., the transfer functions, providing clearer physical interpretation and enabling seamless integration with existing impedance-based small-signal analysis frameworks.


The main contributions of this paper are summarised as follows:
\begin{itemize}
    \item[1)] A modular PINN architecture is developed, with separate network modules dedicated to predicting the poles and residues of the whole-system impedance model in a multi-inverter power system.

    \item[2)] A sophisticated loss function and a training strategy are developed for the PINN so that it can be trained effectively with SRD.
    
    \item[3)] A Sobol-sequence-based sampling strategy is adopted to generate a training dataset with sufficient and uniform coverage of the high-dimensional OP space, enabling efficient training with a limited number of samples.
    
    \item[4)] The proposed PINN is comprehensively validated in a 2-IBR system and a 4-IBR system, with unique useful insights into system stability provided, such as the eigenvalue loci under the variations of power flow and the possible mode range under a specific operating condition. These results are difficult to obtain using conventional methods, whereas they can be readily generated by the PINN in seconds.
\end{itemize}

The rest of this paper is organised as follows. Section~II presents an overview of impedance modelling for multi-inverter MIMO systems and describes the construction of the training data based on Sobol-sequence OP sets and step responses. Section~III details the proposed PINN architecture and the corresponding training procedure. Section~IV presents case studies on multi-inverter MIMO systems to validate the proposed method. Section~V concludes the paper.

\section{Whole-System Impedance Model and Data Preparation}
\subsection{Whole-System Impedance Model}
The whole-system impedance model $\Zsys$, or equivalently, whole-system admittance model $\Ysys$, is a type of frequency-domain networked impedance model represented in a matrix format \cite{Zhu2022RootCause}. For the sake of brevity, this paper discusses only the whole-system admittance model $\Ysys$, and represents it simply as $\Ys$ for brevity. $\Ys$ is a closed-loop model describing the relationship between the small-signal voltage and current across all buses. For an $N$-bus system, we have
\begin{equation} \label{eq_Ysys1}
    \Ys\cdot \Tilde{v}=\Delta i,
\end{equation}
where $\Tilde{v}=[v_1, v_2, \cdots, v_N]^T$ is the input vector of voltage perturbations introduced at each bus, and $\Delta i=[\Delta i_1, \Delta i_2, \cdots \Delta i_N]^T$ is the output vector of current responses at all buses. Due to a three-phase system, $\Ys$ is essentially expressed in a global $\dq$ frame. For each entry of $\Ys$, e.g., the $i$-th diagonal entry $\Ys_{i}$, it is a $2\times 2$ matrix block in $\dq$ frame, such that
\begin{equation}\label{eq_matrix}
    \Ys_i = \left[ \begin{matrix}
    Y_{i,dd}(s) & Y_{i,dq}(s)\\
    Y_{i,qd}(s) & Y_{i,qq}(s)\\
\end{matrix} \right].
\end{equation}

As a frequency-domain model, each element of $\Ys$ is essentially a transfer function and can be represented in a pole-residue format. For example,
\begin{equation}\label{eq_Y_ic}
Y_{i,c}(s)
= \sum_{k=1}^{n} \frac{r_k^{(i,c)}}{s - \lambda_k},
\end{equation}
where $c$ refers to a frame channel, i.e., $c\in \{dd, dq, qd, qq\}$.
Therefore, it is clear that $\Ys$ is a $2N\times 2N$ transfer function matrix. It is demonstrated in \cite{Gu2021Impedance} that all elements of $\Ys$ share the same set of poles $\{\lambda_1, \lambda_2, \cdots \lambda_n\}$, which are the system eigenvalues, i.e., oscillatory modes. The residue of the diagonal element $Y^\text{sys}_{i}$ corresponding to a mode $\lambda_k$ refers to the impedance participation factor, such that 
\begin{equation} \label{eq_DeltaLambda}
\Delta \lambda = \langle - \text{Res}^{*}_\lambda \Ysys_{kk}, \Delta Z_{\text{A}k} \rangle,
\end{equation}
As demonstrated in \cite{Zhu2022RootCause, Zhu2024IMR}, the successful identification of the diagonal elements of $\Ys$ is very useful for small-signal analysis and oscillation early warning in multi-inverter power systems.

\subsection{Definition of OP Space}
The OP of a multi-inverter power system characterises the admissible steady-state operating conditions at each bus. Specifically, for bus-$i$, it contains voltage magnitude $V_i$, phase angle $\theta_i$, active power generation $P_{i}$, reactive power generation $P_{i}$, active load $P_{li}$, reactive load $Q_{li}$, and frequency $F_i$. Since the steady-state frequency $F_i$ is typically fixed and identical across all buses, it is not explicitly considered in the subsequent analysis. The physical quantities $(V_i, \theta_i, P_{i}, P_{i}, P_{li}, Q_{li})$ are determined as the solution of the power flow. Consequently, although the operating point at each bus is characterised by six quantities, they are not all independent; rather, the available degrees of freedom are constrained by the power flow equations and depend on the intrinsic type of the bus. To reduce redundancy, the OP space is defined with only the independent quantities, namely those that can be directly adjusted during system operations. It is worth noting that the network topology is considered unchanged in this work.

We first define the operating space for local buses and loads. For a voltage-type bus, e.g., the infinite bus, or a grid-forming (GFM) inverter which sets both voltage and grid frequency, its local OP space $\OPS_i$ is defined simply as
\begin{equation}
\mathcal{O}_{V,i}=\left\{V_i \in \mathbb{R}\;\middle|\;
V_{i,\min} \le V_i \le V_{i,\max} \right\}.
\end{equation}
Here, the phase angle of the bus is assumed to be fixed for simplicity. 

For a \textit{PV} type bus, e.g., a bus with a GFM inverter, or with a grid-following (GFL) inverter that also has a voltage outer loop, its operating space is defined as
\begin{equation}
\begin{aligned}
\mathcal{O}_{PV,i}
= \Big\{
(V_i, P_{i}) \in \mathbb{R}^2
\;\Big|\;
& V_{i,\min} \le V_i \le V_{i,\max}, \\
& P_{i,\min} \le P_{i} \le P_{i,\max}
\Big\}.
\end{aligned}
\end{equation}

For a \textit{PQ} type bus, e.g., a conventional GFL inverter without a voltage outer loop, its operating space is defined as
\begin{equation}
\begin{aligned}
\mathcal{O}_{PQ,i}
= \Big\{
(P_{i}, Q_{i}) \in \mathbb{R}^2
\;\Big|\;
& P_{i,\min} \le P_{i} \le P_{i,\max}, \\
& Q_{i,\min} \le Q_{i} \le Q_{i,\max}
\Big\}.
\end{aligned}
\end{equation}

For each load, an assumption is made that the load is purely active, and its operating space is defined as
\begin{equation}
\mathcal{O}_{\text{Load},i}=\left\{P_{Li} \in \mathbb{R}\;\middle|\;
P_{Li,\min} \le P_{Li} \le P_{Li,\max} \right\}.
\end{equation}
Here, only active power loads are considered for simplicity.

And eventually, the full OP space $\OPS$ of the multi-inverter system is defined as the Cartesian product of the local operating spaces, such that
\begin{equation}
\begin{aligned}
\OPS &=
\prod_{i \in {I}_V} \OPS_{V,i}
 \times 
\prod_{i \in {I}_{PV}} \OPS_{PV,i}
 \times 
\prod_{i \in {I}_{PQ}} \OPS_{PQ,i}
 \times 
\prod_{i \in {I}_{\text{Load}}} \OPS_{\text{Load},i}, \\
\OPS &\subset \mathbb{R}^{d},
\end{aligned}
\end{equation}
where ${I}_V$, ${I}_{PV}$, ${I}_{PQ}$, denote the index sets of voltage-type \textit{PV} type, and \textit{PQ} type buses, respectively, and ${I}_{\text{Load}}$ denotes the index sets of loads. $d$ refers to the total dimension of the OP space, which is the sum of the dimensions of all local operating spaces. 

The target of the PINN is to have an accurate prediction of the diagonal elements of $\Ys$ at any  $\text{OP}\in \OPS$.

\subsection{Sobol-Based OP Sampling Strategy}
A dataset, which contains system response data measured in an OP set $S_{\text{OP}}$, is essential for training and validation of the PINN. For $S_{\text{OP}}$, we have 
\begin{equation}
   S_{\text{OP}}=\{\mathbf{x}^{(k)}\in\OPS \mid k=1,2,\cdots, N_s\},
\end{equation}
where $\mathbf{x}^{(k)}$ denotes the $k$-th selected OP vector and $N_s$ is the number of samples. To ensure accurate prediction across the OP space, the set $S_{\text{OP}}$ must have strong space-filling coverage of $\OPS$. Therefore, a suitable OP sampling strategy should first be designed to determine $S_{\text{OP}}$.

Uniform sampling is the simplest approach, achieved by applying a uniform step size across all dimensions of the $\OPS$. This approach has been successfully applied in \cite{Liao2024Neural} to sample from a three-dimensional space ($d=3$), resulting in a set of 823 OPs, and in \cite{Wu2025Impedance} to sample from $d=5$ space with 14406 OPs. As clearly seen, such an approach is impractical in an even higher-dimensional OP space, which is normal for a multi-inverter system, because the number of required samples grows exponentially with the number of dimensions in order to maintain comparable coverage. The same situation applies to uniform random sampling. Such a limitation motivates the adoption of a more efficient sampling strategy for high-dimensional OP spaces.


In this work, a low-discrepancy Sobol sequence is employed to generate an $S_{\text{OP}}$ with a good coverage of $\OPS$ under a given sampling budget. $\mathbf{S}_{op}$ is 
Let $\boldsymbol{\xi}^{(k)}=(\xi^{(k)}_1,\cdots,\xi^{(k)}_d)\in[0,1]^d$
denote the $k$-th Sobol point, which represents the normalised coordinate of a sampling point in the OP space. The sample index $k$ admits a binary expansion:
\begin{equation}
    k=\sum_{\ell=0}^{w-1} k_{l+1}\,2^{l},\qquad k_\ell\in\{0,1\},
\end{equation}
where $w=log_2 N_s$ is the number of binary digits required. The $j$-th coordinate of $\boldsymbol{\xi}^{(k)}$ is constructed as
\begin{equation}
\xi^{(k)}_j
=
k_1 v_1^{(j)} \oplus k_2 v_2^{(j)} \oplus \cdots \oplus k_w v_w^{(j)},
\,\, j=1,\cdots,d,
\end{equation}
where $\oplus$ denotes the bitwise XOR operation and
$\{v_i^{(j)}\}$ are predefined direction numbers for dimension $j$.
The Sobol points are mapped to the physical operating space via
\begin{equation}
\mathbf{x}^{(k)}=\Phi\left(\boldsymbol{\xi}^{(k)}\right), \mathbf{x}^{k}\in\OPS,
\end{equation}
where $\Phi$ represents the rescaling process that maps each normalised coordinate to its corresponding physical operating bounds.


This construction yields an OP sampling set with low discrepancy and strong space-filling properties. Under a given sampling budget, the resulting OP set typically provides more uniform coverage of the OP space than equal-step grids or simple random sampling. The corresponding results of the Sobol-based sampling strategy are demonstrated in the case studies. 

\subsection{Training Data Collection}


Once $S_{\text{OP}}$ is determined, system response data needs to be collected at each selected OP through EMT simulations. As summarised in Table \ref{tab:method_comparison}, existing methods employ FRD obtained via FS. Because each round of FS requires injections of sinusoidal waveforms at different FPs, it is extreme time-consuming. For example, injecting 3 cycles at 1 Hz requires at least 3~s total simulation time. As reported in \cite{Zhu2023Injection}, scanning 81 FPs in the range of 1-1000~Hz requires a total simulation time around 30 s. The actual time spent could be even longer for large systems. Such an approach is essentially impractical for MIMO systems where measurements need to be performed at each bus, and the size of $S_{\text{OP}}$ is comparably large. 

To this end, SRD is adopted for training the PINN. At a selected OP $\mathbf{x}\in S_{\text{OP}}$, for bus-$i$, a $1\%$ voltage step change is applied sequentially to the
$d$-axis and $q$-axis. The resulting current responses $y_i(t)$ on both the $d$- and $q$-axes are measured at the same bus, with DC bias removed. Such that we have:
\begin{equation} \label{eq_Inv_Laplace}
\mathcal{L}^{-1}\{\frac{0.01}{s}\Ys_i\}= y_i(t)={\bigg[}\begin{matrix}
	y_{i,dd}(t)&		y_{i,dq}(t)\\
	y_{i,qd}(t)&		y_{i,qq}(t)\\
\end{matrix}\mathrm{\bigg]},
\end{equation}
where $\mathcal{L}^{-1}{\cdot}$ denotes the inverse Laplace transform. 

For each step response, a window length of 1~s is recorded with 1000~Hz sampling rate, leading to a total simulation time of 2~s per bus. This data-acquisition process is substantially shorter than that required by conventional FS–based methods. By recording $y(t)$ at all $\mathbf{x}\in S_{\text{OP}}$ and across all buses, the dataset is prepared, which can then be used to train and validate the PINN.

\section{PINN for Impedance Prediction}

\begin{figure*}
    \centering     
    \includegraphics[width=6.5in]{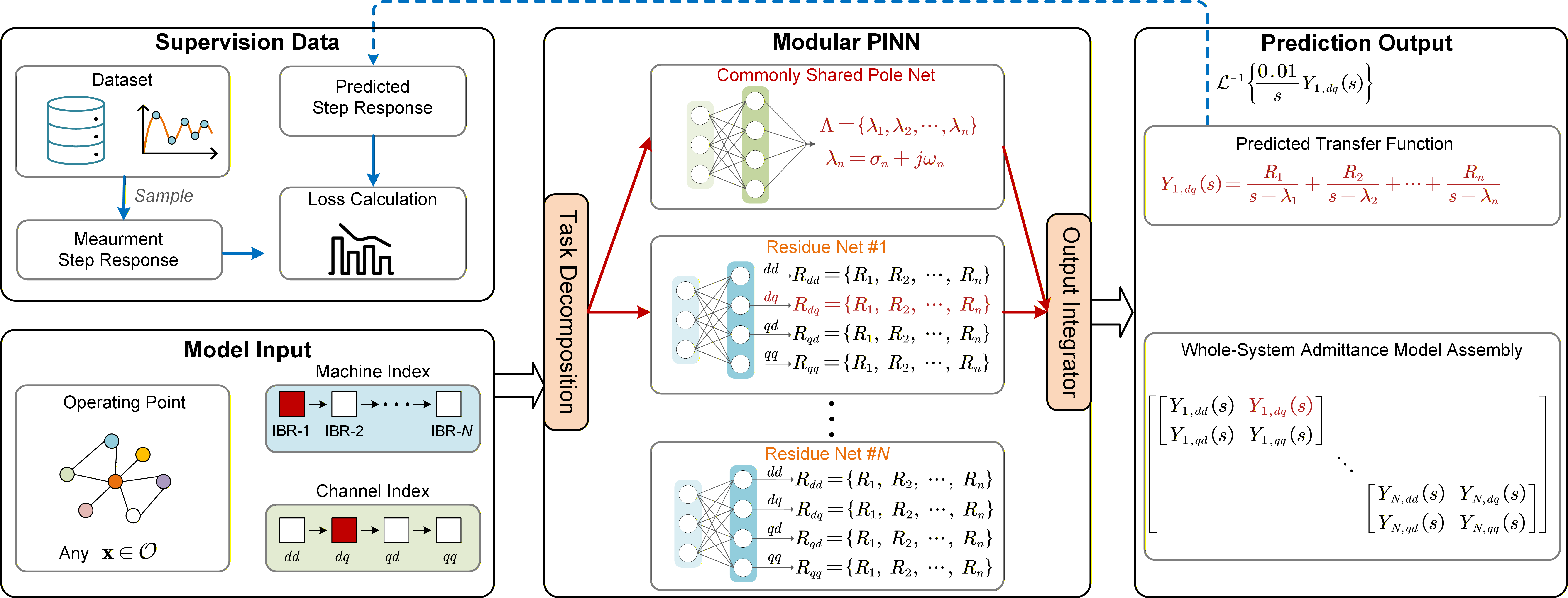} 
    \caption{Architecture of the proposed modular PINN.} 
    \label{ExTF} 
\end{figure*}
\subsection{Learning Task Formulation}
Because the impedance prediction task is formulated as learning a mapping from any $\mathbf{x}\in\OPS$ to the corresponding diagonal elements of the whole-system admittance model $\Ys$,  $\mathbf{x}$ is directly set as the input feature vector. 

To avoid potential optimisation interference from jointly training all IBRs and channels, the dataset is partitioned by IBR and channel indices. 
For the IBR on bus-$i$ and channel $c$, let $y_{i,c}(t)$ denote the corresponding step-response. Each training sample consists of the OP vector $\mathbf{x}$, the index pair $(i, c)$ that uniquely identifies the IBR and channel, and the measured response ${y}_{i,c}$. During training, each forward pass is performed for a specific OP $\mathbf{x}$ and index pair $(i, c)$.

Since the theoretical impedance model is often not directly available, an indirect supervision strategy based on step responses is adopted. Given an OP vector $\mathbf{x}$ and an index pair $(i, c)$, the network outputs an impedance model for IBR $i$ and channel $c$. Let $\hat{Y}_{i,c}(s)$ denote the predicted transfer function for this channel, which is given by
\begin{equation}
    \hat{Y}_{i,c}(s) = F_{\Theta}(\mathbf{x}, i, c),
\end{equation}
where $\Theta$ denotes the PINN parameters. The corresponding step response $\hat{y}_{i,c}$ is then obtained analytically from $\hat{Y}_{i,c}(s)$ using \eqref{eq_Inv_Laplace} and compared with the measured response $y_{i,c}$ to form the training loss.

A selection of system order is also required, which determines the order of the PINN output transfer functions. Although the order of analytical model is usually very high, nearly 100 in the 4-IBR case in this work, most of them are unimportant as they are either far from the imaginary axis or at the frequency over 1000~Hz. To avoid over-fitting issues and to learn only the critical modes, an order number of 20 is selected in the case study. Such a number can also be tuned based on the learning outcomes.

\subsection{Network Architecture}
To learn the OP-dependent impedance model in the pole–residue form of \eqref{eq_Y_ic}, a PINN-based architecture is constructed, as illustrated in Fig.~\ref{ExTF}, the model consists of two components: a pole net that generates system-wide shared poles, and residue nets that produces IBR- and channel-specific residues. Together, these two components constitute the transfer function representation of the impedance model.

\subsubsection{Pole Net}
For a given OP, all IBRs and channels in the system share the same set of poles \cite{Gu2021Impedance}, so a single pole net is used for all IBRs and channels. Given that the physical poles of the system are either real or appear in complex-conjugate pairs, the pole net is designed to generate only poles in the upper half of the complex plane. This prior constraint avoids generating redundant conjugate poles and effectively narrows the search space, which is beneficial for training. The pole net outputs the real and imaginary parts separately, so that each pole is represented as
\begin{equation}
    p_k = \Re\!\left(p_k\right) + j\,\Im\!\left(p_k\right).
    \vspace{-0.2cm}
\end{equation}
During output integration, the poles in the upper half-plane are mirrored with respect to the real axis to form conjugate pairs, which are then used to construct the transfer function and compute the step responses for loss evaluation. In this way, the shared pole set captures the global modal characteristics of the system across all IBRs and channels.

\subsubsection{Residue Net}
The residue net generates the residues associated with the shared poles produced by the pole net. In contrast to the globally shared pole net, the residue net adopts a local structure: an independent residue subnetwork is assigned to each IBR, and the index $i$ is used to select the corresponding subnetwork. For a given index pair $(i,c)$, the residue subnetwork of IBR $i$ takes as input the OP vector $\mathbf{x}$ together with a one-hot encoding of the channel index $c$ and outputs a set of residues
\begin{equation}
    \{ r_k^{(i,c)} \}_{k=1}^{n}.
\end{equation}
These residues are in one-to-one correspondence with the shared poles $\{p_k\}_{k=1}^{n}$ in the pole--residue representation of the impedance transfer function in \eqref{eq_Y_ic}.

This design separates the roles of the two components: the pole net captures the OP-dependent global oscillatory modes through the shared poles, while the residue net determines how these modes are characterized for each IBR and channel. The resulting residues, together with the shared poles, construct the impedance transfer function.

Based on the transfer function, the step response can be acquired from the inverse Laplace transform, which can be analytically computed as
\begin{equation}
\hat{y}_{i,c}(t) = \sum_{k=1}^{n} 2\,\Re\!\left\{ r_k^{(i,c)} \frac{e^{p_k t}-1}{p_k} \right\}.
\end{equation}
In numerical implementation, a small positive constant is added to the denominator to avoid numerical issues when $|p_k|$ is close to zero. The predicted response $\hat{y}_{i,c}(t)$ is then compared with the measured response $y_{i,c}(t)$, and the discrepancy between them is quantified by the loss function for network update. This entire process is implemented in a differentiable manner within the network, enabling end-to-end training.

\subsection{Loss Function and Training Strategy}
Given the analytically computed step responses described in the previous subsection, a sophisticated loss function for supervised training is defined in this subsection. For a given sample $(\mathbf{x},i,c)$, let the measured and predicted step responses, which are sampled on a common time grid, be denoted by $y_{i,c}[n]$ and $\hat{y}_{i,c}[n]$, $n = 0,\ldots,T_n-1$. The loss function is designed to quantify the discrepancy between these sequences while capturing the key characteristics of the step response and accommodating the multi-converter, multi-channel setting.

To mitigate scale differences across IBRs and channels, the peak-to-peak amplitude of the ground-truth sequence is used as a normalization factor. Specifically, the maximum and minimum values of the measured sequence are defined as
\begin{equation}
y_{\max} = \max_n\, y_{i,c}[n], \qquad
y_{\min} = \min_n\, y_{i,c}[n].
\end{equation}
and the normalization factor is given by
\begin{equation}
\Delta_{i,c} = y_{\max} - y_{\min} + \varepsilon, \qquad \varepsilon > 0,
\end{equation}
where $\varepsilon$ is a small positive constant to prevent division by zero. Both the measured and predicted sequences are then normalized by this factor:
\begin{equation}
\tilde{y}_{i,c}[n] = \frac{y_{i,c}[n]}{\Delta_{i,c}}, \qquad
\tilde{\hat{y}}_{i,c}[n] = \frac{\hat{y}_{i,c}[n]}{\Delta_{i,c}}.
\end{equation}

Most discriminative information in a step response is concentrated in the early transient, whereas the steady tail often varies slowly and carries less information on dynamics. If all time samples are weighted equally, the mean-squared error can be dominated by this long tail, reducing sensitivity to dynamics in the transient. To emphasise early dynamics while still constraining steady-state bias, an exponential weight is applied to the normalised time position:
\begin{equation}
r_n = \frac{n}{T_n-1} \in [0,1],
\end{equation}
with
\begin{equation}
w_n = A_we^{-\mu r_n}, \,\, A_w>0,\ \mu>0 .
\end{equation}
Here, $\mu$ controls the decay rate, and $A_w$ is a scaling factor that adjusts the overall weight level. These two parameters can be tuned during the training process.

Under the above normalisation and time weighting, the weighted mean-squared error (WMSE) for a given IBR and channel is defined as
\begin{equation}
\mathcal{L}_{\mathrm{sample}}(\mathbf{x},i,c)
=\frac{1}{T}\sum_{n=0}^{T-1}
w_n
\big(\tilde{\hat{y}}_{i,c}[n]-\tilde{y}_{i,c}[n]\big)^2 ,
\label{eq:per-sample-loss}
\end{equation}
where $\tilde{y}_{i,c}[n]$ and $\tilde{\hat{y}}_{i,c}[n]$ are the normalized measured and predicted responses for the sample $(\mathbf{x},i,c)$.

Let $\mathcal{D}$ denote the training dataset, consisting of samples
$(\mathbf{x}, (i,c), \mathbf{y}_{i,c})$. The model is trained by minimising the
step-response loss over the dataset. The overall training procedure is summarized
in Algorithm~\ref{alg:extfnet-train}. For each training sample, the model performs
shared-pole inference and per-converter, per-channel residue computation, followed
by analytical step-response evaluation and loss calculation. The network parameters
are then updated through an iterative optimisation procedure.
\begin{algorithm}
\caption{Training Pipeline}
\label{alg:extfnet-train}
\begin{algorithmic}[1]
\State Initialize network parameters $\Theta=\{\Theta_{\text{pole}},\Theta_{\text{res}}\}$, stability constant $\varepsilon>0$, and time-weight hyperparameters $A>0$, $\lambda>0$.
\For{each training iteration}
  \State \textit{Prepare the current mini-batch:} 
  \[
     \mathcal{S}=\{(\mathbf{x},(i,c),\mathbf{y})\},
  \]
  \State For each $(i,c)$, build a one-hot indicator vector $\mathbf{e}_{i,c}$.
  \State \textit{Forward pass and sample-wise loss computation:}
  \For{each $(\mathbf{x},(i,c),\mathbf{y})\in\mathcal{S}$}
    \State \textit{PoleNet (shared poles):}
    \State $(\tilde{\mathbf{p}}^{(\Re)},\tilde{\mathbf{p}}^{(\Im)}) \gets \mathrm{PoleNet}_{\Theta_{\text{pole}}}(\mathbf{x})$
    \State $p^{(\Re)} \gets -\mathrm{softplus}(\tilde{\mathbf{p}}^{(\Re)})$, 
    \State $p^{(\Im)} \gets \mathrm{softplus}(\tilde{\mathbf{p}}^{(\Im)})$
    \State $\mathbf{p} \gets p^{(\Re)} + j\,p^{(\Im)} \in \mathbb{C}^{N}$
    \State \textit{ResidueNet (per converter/channel):}
    \State $\mathbf{r} \gets \mathrm{ResidueNet}_{\Theta_{\text{res}}}(\mathbf{x},\,\mathbf{e}_{i,c}) \in \mathbb{C}^{N}$
    \State \textit{Analytical decoder (step response):}
    \State $\mathbf{p}_\varepsilon \gets \mathbf{p} + \varepsilon$
    \State Given the time grid $\mathbf{t}=\{t_n\}_{n=0}^{T-1}$, compute
    \[
      \hat{\mathbf{y}} \gets \sum_{k=1}^{n} 2\,\Re\!\left\{ r_k \frac{e^{p_k\,\mathbf{t}}-1}{p_{\varepsilon,k}} \right\} \in \mathbb{R}^{T}.
    \]
    \State \textit{Peak-to-peak normalization:}
    \State $s \gets (\max_n y_n - \min_n y_n) + \varepsilon$
    \State $\tilde{\mathbf{y}} \gets \mathbf{y}/s$, \quad $\tilde{\hat{\mathbf{y}}} \gets \hat{\mathbf{y}}/s$
    \State \textit{Exponentially time-weighted step-response loss:}
    \State $L_{\mathrm{sample}} \gets 0$
    \For{$n=0,\ldots,T-1$}
      \State $r_n \gets \dfrac{n}{T-1}$,\quad $w_n \gets A\,e^{-\lambda r_n}$
      \State $L_{\mathrm{sample}} \gets L_{\mathrm{sample}} + \dfrac{1}{T}\, w_n \big(\tilde{\hat{y}}_n-\tilde{y}_n\big)^2$
    \EndFor
  \EndFor
  \State \textit{Aggregation over the mini-batch:}
  \State $L \gets \dfrac{1}{|\mathcal{S}|}\,\displaystyle\sum_{(\mathbf{x},(i,c),\mathbf{y})\in\mathcal{S}} L_{\mathrm{sample}}(\mathbf{x},i,c)$
  \State \textit{Parameter update:} $\Theta \leftarrow \mathcal{O}\!\big(\Theta,\,\nabla_{\Theta} L\big)$
\EndFor
\end{algorithmic}
\end{algorithm}

\section{Case Studies}
To demonstrate the effectiveness of the proposed PINN-based whole-system admittance prediction framework on IBR-based systems of different scales, two case studies were conducted. A 2-IBR system was first considered to examine prediction performance, data efficiency, and the ability to capture eigenvalue loci under OP variations. The same evaluation procedure was then applied to a 4-IBR system to investigate the scalability of the proposed method in higher-dimensional OP spaces of larger-scale power systems. Both the analytical model and the EMT model were established in Matlab/Simulink. If not specifically mentioned, all quantities are expressed in per unit (\pu). Given the consistency between the analytical and EMT models, the SRD data used for training were generated directly from the analytical model for the case studies. The neural network models were developed using Python~3.9.18 with PyTorch~2.2.1. All tests were conducted on a workstation equipped with an NVIDIA GeForce RTX~5080 GPU, 64~GB of RAM, and an Intel Core Ultra~9~285K processor.
\subsection{Validation on a 2-IBR System}

\subsubsection{System Description and OP Space}

\begin{figure}
    \centering     \includegraphics[width=0.48\textwidth]{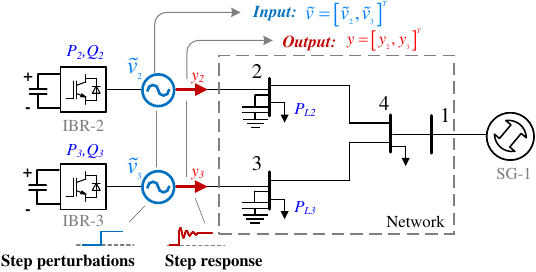} 
    \caption{One-line diagram of the 2-IBR system: the voltage step perturbation is added at each bus while the corresponding current responses are recorded.} 
    \label{F_model_2bus} 
    \vspace{-0.2cm}
\end{figure}

The system consists of two IBRs and one SG, as demonstrated in Fig.\ref{F_model_2bus}. The synchronous generator SG-1 is connected to bus~1. IBR-2 and IBR-3 are connected to buses~2 and~3, respectively. The two IBRs are tuned with different parameters, with IBR-3 being worse-tuned with a lower inner current control bandwidth. Both buses~2 and~3 are classified as \textit{PQ} type buses. A purely active load is connected at each of buses~2 and~3, whose operating variables are the load active powers $P_{L2}$ and $P_{L3}$, respectively. Voltage at bus-1 is considered fixed at 1 \pu for simplicity. Since buses~2 and~3 are \textit{PQ} type buses, their independent operating variables are the active and reactive powers $(P_{i}, Q_{i})$, where $i \in \{2,3\}$.

The operation constrains are given as
\begin{align}
 &P_{2},P_{3} \in [0,0.95],\quad Q_{2},Q_{3} \in [-0.05,0.05] \\
 &P_{L2}, P_{L3}\in[0,0.5].
\end{align}

Accordingly, the OP vector $\mathbf{x}$ of the system is defined as
\begin{equation}
\mathbf{x} =
\big[
P_{2},\;
Q_{2},\;
P_{3},\;
Q_{3},\;
P_{L2},\;
P_{L3}
\big]^{\top}
\in \mathbb{R}^{6}.
\end{equation}

The system maintains stable in the operation range and the objective of the proposed PINN is to predict the diagonal elements of $\Ys$ associated with buses~2 and~3, where the IBRs are connected.

\subsubsection{OP Sampling and Data Generation}
To achieve effective coverage of the OP space under a limited sampling budget, Sobol low-discrepancy sequences are adopted in this study to generate OP samples, owing to their favorable space-filling properties in high-dimensional spaces.

\begin{figure}[!t] 
    \centering     \includegraphics[width=3.2in]{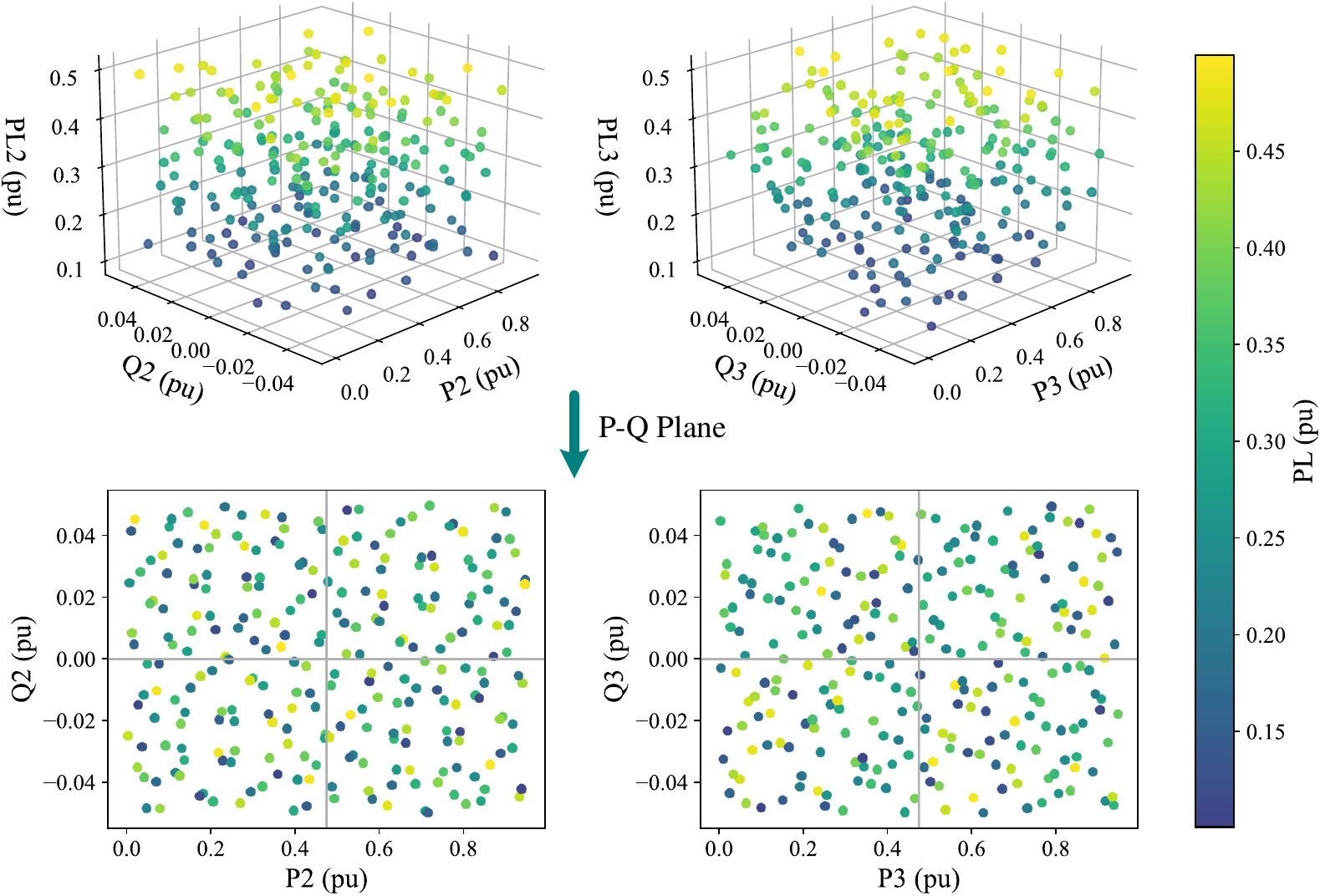} 
    \caption{Distribution of OPs based on Sobol sequence.} 
    \label{F_sobol} 
\end{figure}

Fig. \ref{F_sobol} illustrates the OP distribution generated using the Sobol sequence. The six-dimensional OPs associated with the two IBRs are visualized using two three-dimensional scatter plots, together with their two-dimensional projections onto the $P$--$Q$ planes. The samples are distributed uniformly across different dimensions of the OP space, without pronounced clustering or large coverage gaps, indicating that the adopted sampling strategy provides an effective coverage of the OP space.

To quantitatively assess the uniformity of the OP-space coverage, the $L_2$ discrepancy is adopted as the coverage metric. For a set of $N$ sampling points $P_N=\{\mathbf{x}_i\}_{i=1}^{N} \subset [0,1]^d$, the $L_2$ discrepancy is defined as
\begin{equation}
D_{N,2}(P_N)
=
\left(
\int_{[0,1]^d}
\left|
\Delta(\mathbf{t}; P_N)
\right|^2
\, d\mathbf{t}
\right)^{1/2},
\end{equation}
where $\Delta(\mathbf{t}; P_N)$ is the associated local discrepancy function, and a smaller value of $D_{N,2}$ indicates a more uniform and space-filling distribution of the sampling points.

\begin{figure}[!t] 
    \centering     \includegraphics[width=3.2in]{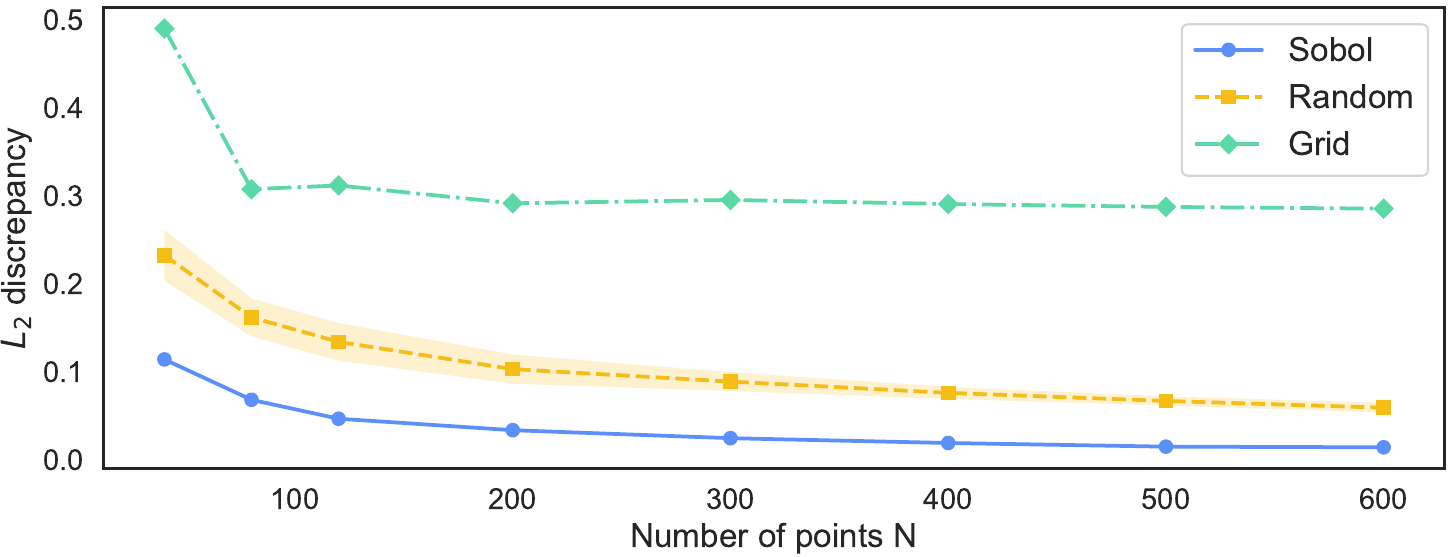} 
    \caption{Comparison of $L_2$ Discrepancy for Different Sampling Strategies.} 
    \label{F_L2} 
\end{figure}

Fig.~\ref{F_L2} compares the $L_2$ discrepancy obtained using Sobol sampling, independent uniformly distributed random sampling, and subsets of uniform grid points. For the same sample size, Sobol sampling consistently achieves lower discrepancy values and a smoother decreasing trend as the number of samples increases. In comparison, random sampling exhibits larger variability, while grid-based subsets result in higher discrepancy levels in high-dimensional spaces under finite sample sizes.
\begin{figure}[!t] 
    \centering     
    \includegraphics[width=0.48\textwidth]{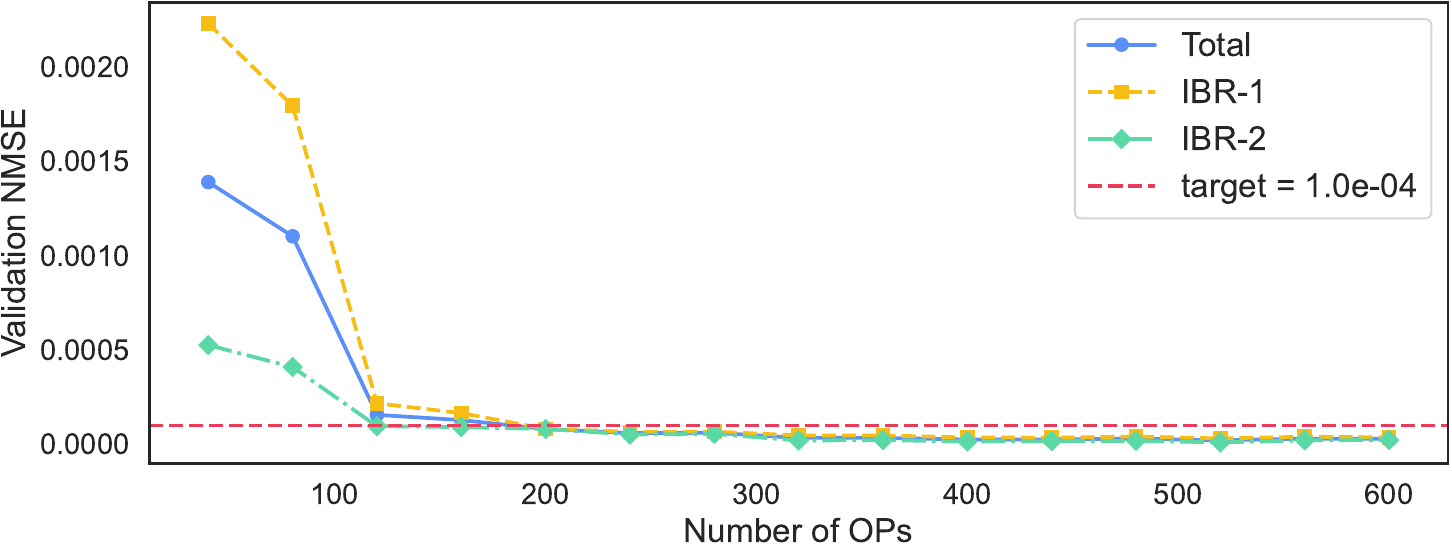} 
    \caption{Scaling of validation NMSE with number of OPs in the 2-IBR system.} 
    \label{F_scaling} 
    \vspace{-0.2cm}
\end{figure}

For each OP, 
a 1~s step-response window, with a sampling rate of 1000~Hz is recorded. This window captures the dominant transient dynamics and oscillatory characteristics of the system. Compared with FS-based approaches, recording only SRD significantly reduces data-collection costs in high-dimensional OP spaces.

To evaluate the impedance prediction performance of the proposed PINN, its data efficiency under limited training data and its prediction performance at unseen operating points are examined. 


Data efficiency is evaluated by varying the number of training OPs from 40 to 600, with a fixed validation set and an independently trained PINN for each training data size. Once the transfer functions are predicted, their SRD can be readily achieved. The average normalised mean squared error (NMSE) between the predicted and simulated SRD associated with the validation OPs is used as the evaluation metric. As shown in Fig.~\ref{F_scaling}, the validation NMSE decreases rapidly as the number of training OPs increases. When the number of training OPs reaches approximately 200, the prediction error falls below the predefined threshold of $1\times10^{-4}$, beyond which further increasing the training data results in only marginal performance improvement. This suggests that reliable impedance prediction can be achieved with a relatively small number of training OPs.

\subsubsection{Data Efficiency and Prediction Performance}

\begin{figure*}[t] 
    \centering     
    \includegraphics[width=6.3in]{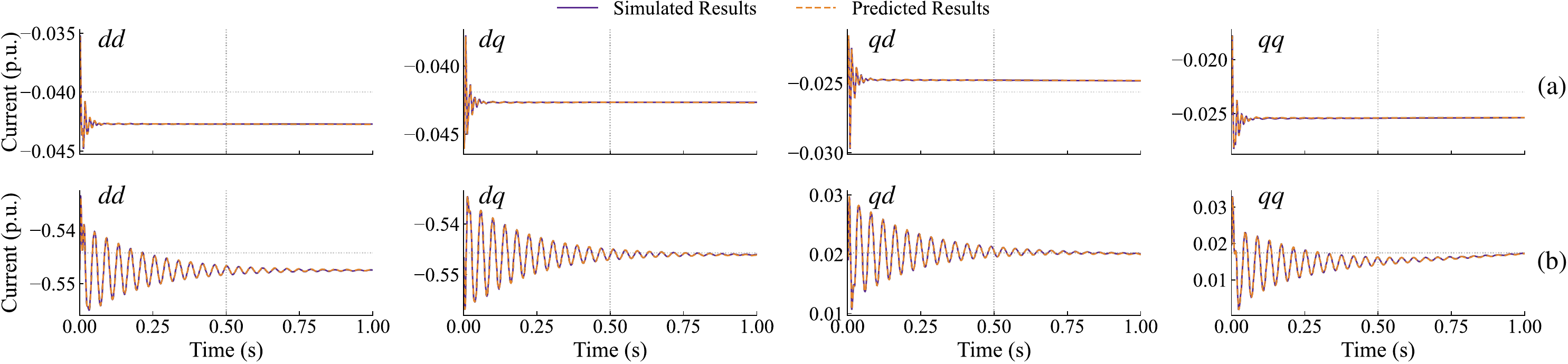} 
    \vspace{-0.25em}
    \caption{Predicted and simulated step responses at a representative unseen operating point. (a) IBR--2, (b) IBR--3.} 
    \vspace{-0.25em}
    \label{F_step} 
\end{figure*}

Based on the above data-efficiency results, a representative unseen operating point is selected from the validation set to further assess the prediction accuracy. At the selected unseen operating point, the predicted step responses of both IBRs in all four \(dq\)-frame channels show good agreement with the EMT simulation results, as illustrated in Fig.~\ref{F_step}. The dominant oscillatory modes and the main transient dynamics of the system are well reproduced, including the oscillation frequencies, damping characteristics, and steady-state responses.

\begin{figure*}[t] 
    \centering     
    \includegraphics[width=6.3in]{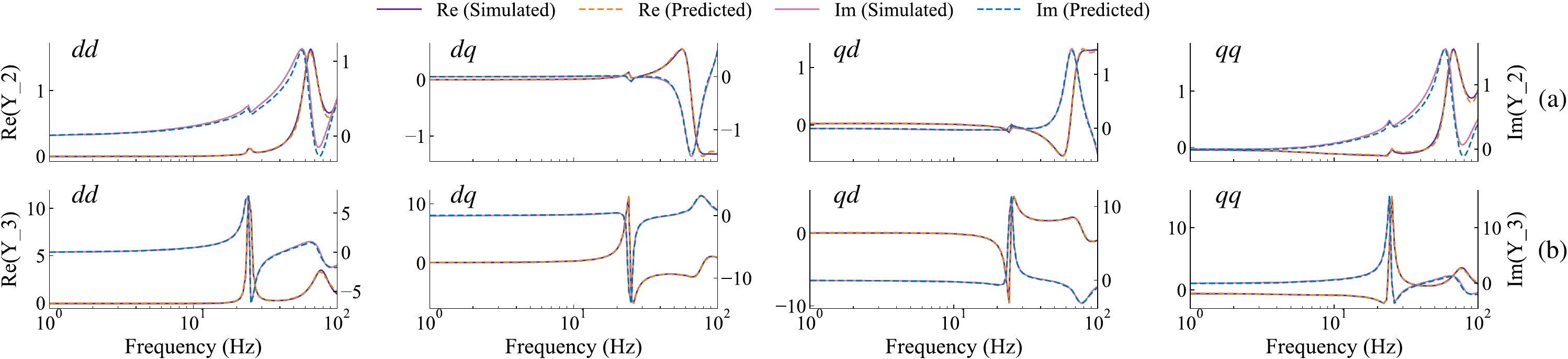} 
    \vspace{-0.25em}
    \caption{Predicted and simulated frequency responses at a representative unseen operating point. (a) IBR--2, (b) IBR--3.} 
    \vspace{-0.25em}
    \label{F_complex} 
\end{figure*}

Building on the above results, the frequency responses reconstructed from the predicted pole–residue representations are further compared with the simulation results, as shown in Fig.~\ref{F_complex}. Good agreement is observed over the frequency range of interest. The prediction accuracy for IBR-1 is slightly lower than that for IBR-2, which is consistent with the higher participation of IBR-2 in the dominant oscillatory modes under the considered system configuration. Despite this difference, the key resonance features and dominant dynamic behaviors are captured for both IBRs. These results suggest that the proposed PINN has learnt the underlying dynamic characteristics of this 2-IBR system, leading to consistent prediction performance in both the time and frequency domains.

\subsubsection{Eigenvalue Locus Analysis under OP Variations}
The eigenvalue loci under OP variations provide useful insights into the evolution of the system's dynamics with power flow. While such loci can only be produced through a large number of EMT simulations with conventional approaches, they are readily available from the trained PINN and can be generated in seconds. 

\begin{figure} [t]
    \centering     \includegraphics[width=3in]{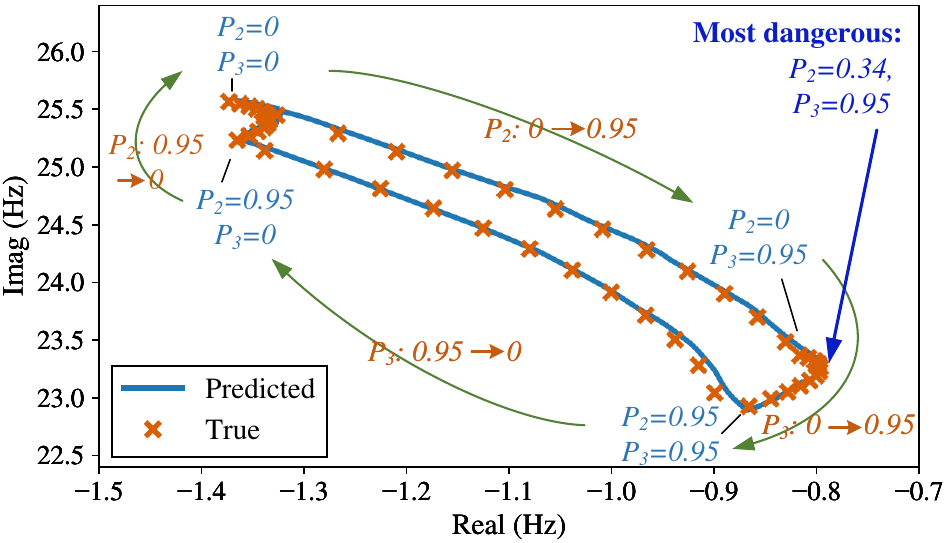} 
    \caption{Locus of the dominant oscillatory mode subject to a sequence of OP variation in the 2-IBR system: a unique view provided by the proposed PINN.} 
    \label{F_pole} 
\end{figure}

In this case study, the dominant oscillatory mode is referred to as Mode--1. A sequence of OP variation is created for demonstration: \(P_2\) and \(P_3\) initially started from zero, \(P_3\) was first increased to 0.95, followed by an increase of \(P_2\) to 0.95, after which \(P_3\) and \(P_2\) were back to zero sequentially. Fig.~\ref{F_pole} shows the eigenvalue locus of Mode--1 from the prediction of PINN and the true modes from the analytical model at some OPs on this sequence. It is evident that the predicted eigenvalue locus exhibits excellent agreement with the true locus and captures the overall evolution of the eigenvalues accurately. Another observation is that near \(P_2 = 0.34\) and \(P_3 = 0.95\), the dominant eigenvalue approaches the imaginary axis most closely, corresponding to the most dangerous operating scenario. This observation contradicts the traditional expectation that the worst case occurs when both \(P_2\) and \(P_3\) reach their maximum values, indicating that experience-based judgement can be unreliable in dynamic studies while the PINN-based approach shows certain advancement.

\subsection{Validation on a 4-IBR System}

\subsubsection{System Description and OP Space}

\begin{figure}[t] 
    \centering     \includegraphics[width=0.49\textwidth]{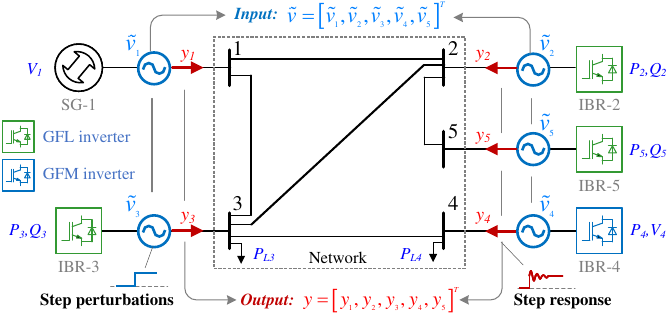} 
    \caption{One-line diagram of the 4-IBR system: the voltage step perturbation is added at each bus while the corresponding current responses are recorded.} 
    \label{F_model_5bus}
    \vspace{-0.4cm}
\end{figure}

The system structure is illustrated in Fig.~\ref{F_model_5bus} and consists of four IBRs and one SG. SG-1 is connected to bus~1. IBR-2, IBR-3, and IBR-5 are GFL IBRs with different parameters connected to buses~2,~3, and~5, respectively. IBR-4 is a GFM IBR with droop controller connected to bus~4. Buses~2,~3, and~5 are classified as \textit{PQ} type buses, hence their independent operating variables are $(P_{i}, Q_{i})$. The independent operating variable associated with bus~1 is the voltage magnitude $V_1$, while for bus~4 they are $(V_4, P_4, )$. Loads at bus~3 and bus~4 are considered varying while other loads are fixed, such that $(P_{L3},P_{L4})$ are independent variables.

The constraints applied in this case are
\begin{equation}
\begin{aligned}
&P_{i} \in [0,0.95],\quad & Q_{i} \in [-0.05,0.05], && i\in\{2,3,5\},\\
&V_1 \in [0.95,1.05],\quad &V_4 \in [1.02,1.05],\\
&P_{4} \in [0,0.5],\quad &P_{L3},P_{L4} \in [0,0.1].
\end{aligned}
\end{equation}
Accordingly, the OP vector $\mathbf{x}$ of the system is defined as
\begin{equation} \label{eq_OP_5bus}
\mathbf{x} =
\big[
V_1,
P_{2},
Q_{2},
P_{3},
Q_{3},
V_{4},
P_{4},
P_{5},
Q_{5},
P_{L3},
P_{L4},
\big]^{\top} \in \mathbb{R}^{11}.
\end{equation}

The objective of the proposed PINN is to predict the diagonal elements of $\Ys$ associated with buses~1,~2,~3,~4, and~5.

\subsubsection{Data Efficiency and Prediction Performance}
In the 4-IBR system, the dimensionality of the OP space is higher than that in the 2-IBR case, making impedance prediction under limited training data more challenging. To evaluate the data efficiency of the proposed method in this higher-dimensional OP space, the same training and validation strategy as in the previous case study is adopted, and the prediction performance is examined under different numbers of training OPs.

\begin{figure}[t] 
    \centering     
    \includegraphics[width=3in]{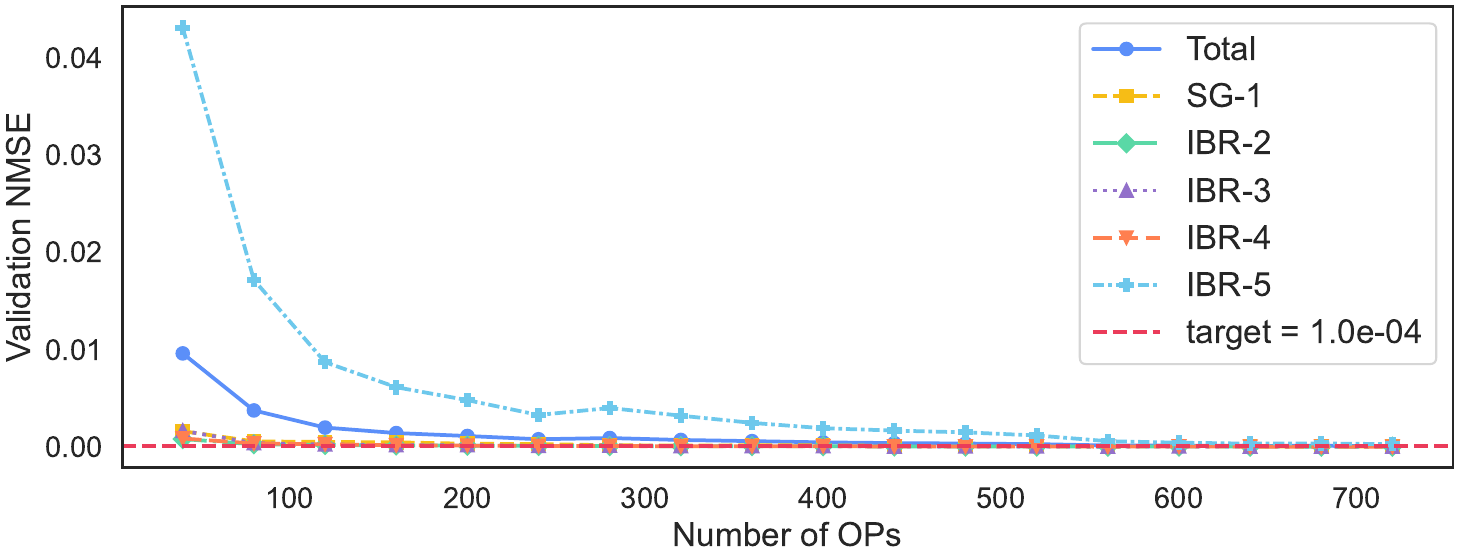} 
    \caption{Scaling of validation NMSE with number of OPs in the 4-IBR system.} 
    \label{F_scaling_5bus} 
\end{figure}

As shown in Fig. \ref{F_scaling_5bus}, the prediction error on the validation set decreases as the number of training OPs increases. When approximately 600 OPs are used for training, the prediction error reaches the predefined accuracy threshold. Despite the increased OP dimensionality, the proposed method is able to achieve acceptable prediction accuracy with a limited increase in the number of training OPs, without requiring an exponential growth in training data to maintain comparable performance.

\begin{figure}[t] 
    \centering     
    \includegraphics[width=0.5\textwidth]{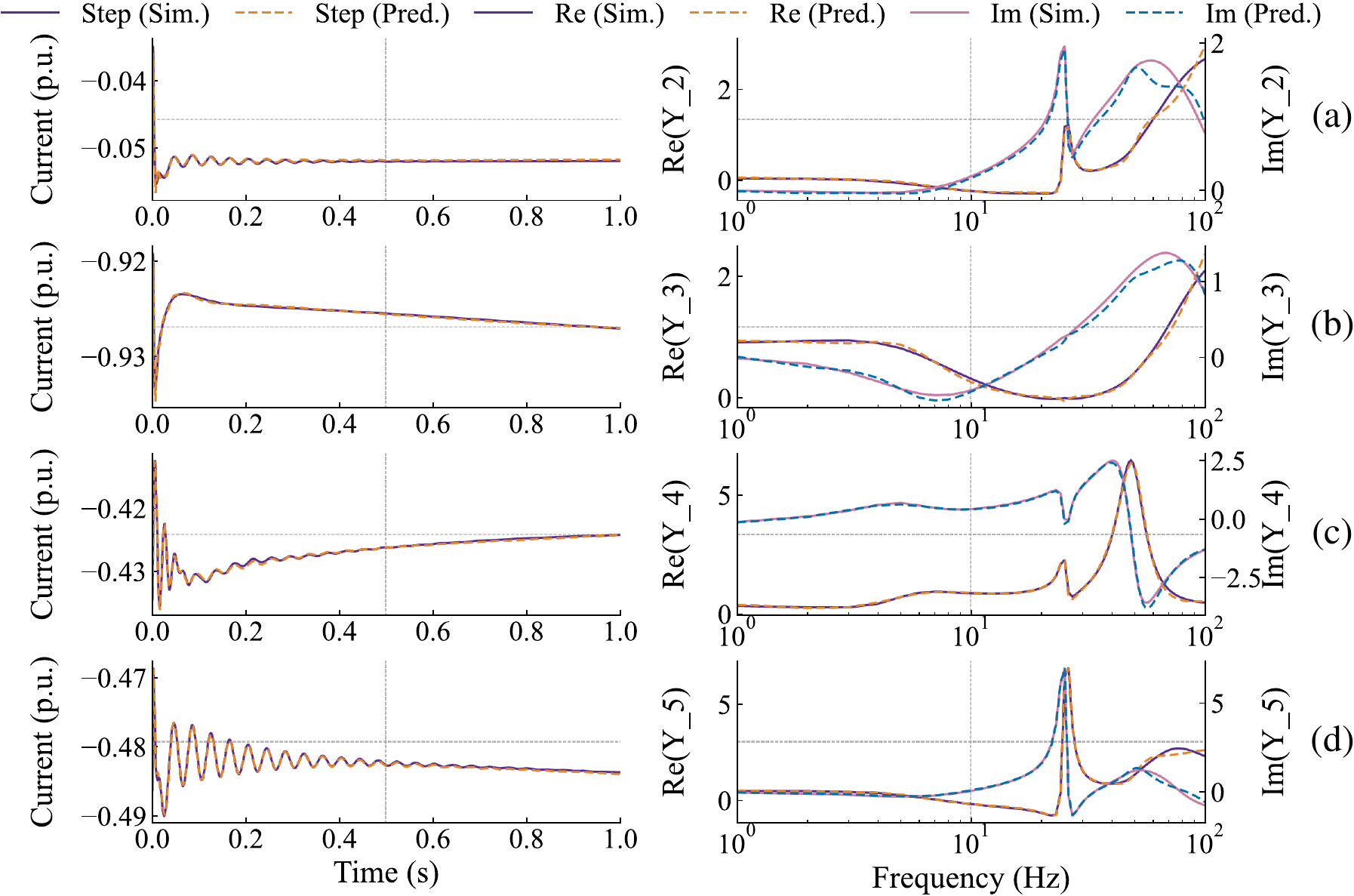} 
    \vspace{-0.25em}
    \caption{Predicted and simulated step (left) and frequency (right) responses at a representative unseen operating point: (a)–(d) IBR–2 to IBR–5, $dd$ channel.} 
    \vspace{-0.25em}
    \label{F_mix_bus5} 
\end{figure}

To further examine the prediction performance, a representative unseen OP is selected from the validation set. As illustrated in Fig. \ref{F_mix_bus5} left, the predicted step responses at this OP show good agreement with reference, demonstrating the accuracy of predictions. Further, the corresponding frequency-response results are shown in Fig. \ref{F_mix_bus5} right. Over the frequency range of interest, the predicted responses exhibit consistent trends, therefore can accurately predict the dynamics of the entire system. 



\subsubsection{Stability-Constrained Power Planning}
\begin{figure}[t] 
    \centering     
    \includegraphics[width=0.4\textwidth]{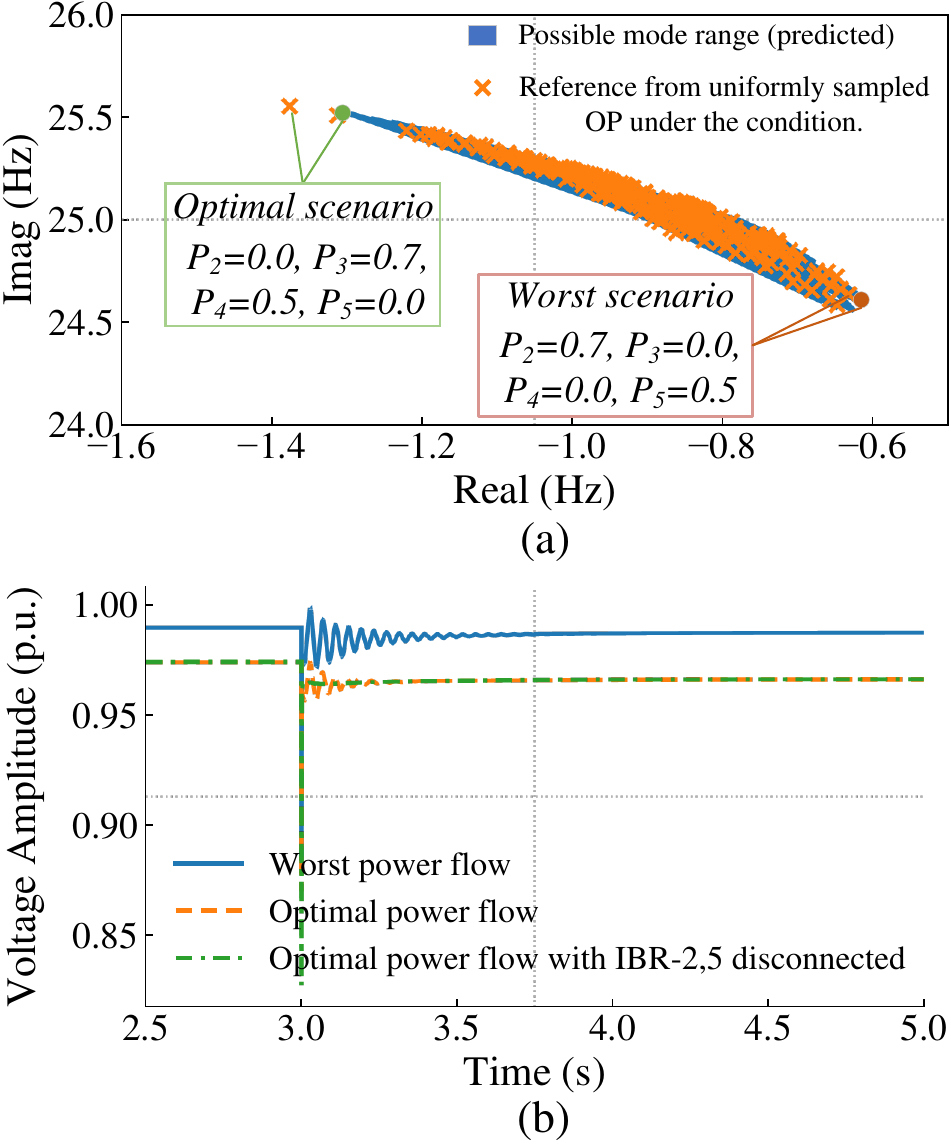} 
    \vspace{-0.25em}
    \caption{(a) Predicted mode range versus reference values under the power flow condition in \eqref{eq_PF_limit}, showing accurate predictions. (b) EMT simulation of the voltage amplitude at bus 2 following a 20\% load step applied to the same bus: the worst-case power flow exhibits pronounced oscillatory behaviour, whereas the optimal scenario shows an improved response. When both IBR-2 and IBR-5 are disconnected (due to zero power output), the oscillations are further damped. The EMT results are consistent with the prediction of the proposed PINN.} 
    \vspace{-0.3cm}
    \label{F_scope} 
\end{figure}

A proof of concept is also presented to demonstrate the use of the proposed PINN for stability-constrained optimal power planning. Specifically, the four IBRs are scheduled to supply a total active power of 1.2 \pu, with the output power of each GFL IBR constrained by a cap of 0.7 \pu., such that:
\begin{equation} \label{eq_PF_limit}
    P_{2}+P_{3}+P_{4}+P_{5}\equiv 1.2, \,\,\,\, P_{2}, P_{3}, P_{5}\in [0,0.7].
\end{equation}
With the proposed PINN, the range of the dominant oscillatory mode under this generation condition can be readily depicted, as shown in Fig. \ref{F_scope} (a). It is apparent that the worst scenario arises when $\{P_{2}=0.7, P_{3}=0, P_{4}=0, P_{5}=0.5\}$, while $\{P_{2}=0, P_{3}=0.7, P_{4}=0.5, P_{5}=0\}$ is an optimal solution. This aligns with the fact that IBR-2, IBR-5 have deliberately detuned parameters. EMT simulations results demonstrated in Fig. \ref{F_scope} (b) further proves the predictions of PINN. This application scenario demonstrates the potential of the proposed PINN in optimal generation planning, realising the generation capability of renewables while maintaining system stability.

\section{Conclusion}
This paper proposes a modular PINN for small-signal analysis in high-dimensional multi-inverter power systems. 
By employing a sobol-sequence strategy to generate OP samples with good space coverage and training the PINN using step response data, the proposed method is able to directly predict the poles and residues of the whole-system impedance model with greatly reduced data acquisition effort. For the PINN, a sophisticated loss function with a time-dependent weighted coefficient is also designed. Validation results on both 2-IBR and 4-IBR systems demonstrate that the proposed approach maintains excellent data efficiency and high prediction accuracy in multi-inverter power systems, and is able to capture the evolution of oscillatory modes under OP variations and predict possible mode range under certain power flow conditions.
\vspace{-0.3cm}

\bibliographystyle{IEEEtran}
\small\bibliography{mybib}

\end{document}